# Energy-Exergy Analysis and Optimal Design of a Hydrogen Turbofan Engine

Mohammadreza Sabzehali[1], Somayeh Davoodabadi Farahani[1], Amir Mosavi[2,3]

[1] *School of Mechanical Engineering, Arak University of Technology, 38181-41167, Arak, Iran*

[2] *John von Neumann Faculty of Informatics, Obuda University, 1034 Budapest, Hungary*

[3] *Institute of Information Engineering, Automation and Mathematics, Slovak University of Technology, 812 37 Bratislava, Slovakia*

**Abstract**

In this study, the effect of inlet air cooling and fuel type on the performance parameters of thrust-specific fuel consumption (TSFC), thermal and exergetic efficiencies, entropy generation rate, and Nitrogen oxide emission intensity index (SNOx) of the GENX 1B70 engine is analyzed in two states of take-off and on design. The results show that with a 20˚C reduction in inlet air temperature on design conditions and JP10 fuel usage, the thermal efficiency and entropy generation rate, thrust and fuel mass flow rate, and TSFC of the engine increase by 1.85, 16.51%, 11.76%, 10.53%, and 2.15% and SNOx & exergetic efficiency decrease by 2.11% and 26.60%, respectively. Also, optimization of the GENX 1B70 engine cycle as hydrogen fuel usage with three separate objective functions: thrust maximization, thermal efficiency maximization, and propulsive efficiency maximization on design point condition was performed based on the Genetic algorithm. Based on the economic approach and exero-environmental, the best cycles from the optimal states were selected using the TOPSIS algorithm. In on design conditions, entropy generation rate, nitrogen oxide production rate, and TSFC for the chosen cycle based on economic approach +18.89%, +10.01%, and -0.21%, respectively, and based on exero-environmental approach -54.03%, -42.02%, and +21.44% change compared to the base engine, respectively.

## 1-Introduction

Turbofan (TF) engines are one of the widely used types of aero gas turbine engines. The design of aero gas turbine engines is very complex and it needs to analyze the thermodynamic cycle of its. Its main parameters include compression ratio, turbine inlet temperature, and by-pass ratio. Optimizing these parameters is very important for purposes such as increasing thrust force and reducing thrust-specific fuel consumption (TSFC). Turbofan engines are separated into two kinds including mixed-flow turbofan and un-mixed flow turbofan. Also it can also be divided into two parts: hot flow and by-pass flow. Also, at the unmixed flow turbofan engines the output flow of bypass channel and low-pressure turbine exhausted to the environment separately through separate nozzles. But, at the mixed-flow turbofan engine, the output flow of the by-pass channel and the low-pressure turbine are combined in the mixer and then imported to the nozzle[1]. Exergetic and energetic analysis of various types of gas turbine cycles has been performed in previous studies. Ibrahim et al.[2] performed energy investigation and exergetic examination of a aero-GT system. They calculated that chemical exergy and physical exergy are related to the rates of the input and output compressor's exergy flow, combustors, and the turbine. The fallouts disclosed that the highest exergy destruction rate belongs to the combustor. Exergy efficiency and energy efficiency of air compressors are 94.9 and 92%, respectively. Also, the exergy and energy efficiencies of the combustor are 67.5% and 61.8%, respectively, also the exergy and energy efficiencies of the turbine are 92% and 82%, respectively. Gamannossi et al.[3] studied performance analysis and emission analysis of GT26 single shaft gas turbine engine. Their results indicated that with increasing turbine inlet temperature (TIT) and compression ratio SNOx increased. Koc et al.[4] performed a gas turbine cycle exergetic analysis with natural gas as fuel. Their fallouts confirmation is that the maximum thermal and exergetic efficiencies for the cycle are 36.45% and 50.50%, respectively. Zhou et al.[5] analyzed and optimized a 3-shaft engine with a turbine with a variable area nozzle. They proposed the control strategy for turbine shaft speed and VAN angle. Pierezan et al.[6] optimized a heavy-duty gas turbine performance using the Cultural coyote algorithm. As a result of this thrust-specific fuel consumption (TSFC) optimization, TSFC has been reduced by 3.6%. Previous researches like Bontempo and Manna[7], Baakeem et al.[8], and Najjar and Abubaker[9] displayed that lessening the intake-air temperature leads to an increase in the density of intake air. Therefore, the mass flow rate of intake air intensifications. growing the intake mass flow rate leads to changes in some engine performance parameters for example thrust force and thermal efficiency. In a study of Bontempo and Mann[7], they optimized the gas turbine cycles based on adding an intercooler and a reheat. They concluded that the energy efficiency of the cycle includes intercooler and reheat 24.96% higher than simple cycle. In another study, Baakeem et al.[8] examined the improvement of the performance of a gas turbine power plant by inlet air cooling techniques. Their results show that in optimum conditions, the inlet air temperature drop is 8 K cooling, and the cooling capacity is 36 kW. Also, in another similar study. Najjar and Abubaker[9] investigated the effect of inlet air cooling on gas turbine engines with heat recovery and exergetic optimization. The study results indicate that the best value of second law efficiency in the optimization results is 40%.

Due to the chemical composition, the type of fuel affects the performance of the engine, and research has been done in this field. Earlier studies show that in aero-GT engines, the type of fuel is effective in the results of engine exergy analysis. Also, Derakhshandeh et al.[10] simulated the environmental and economical analysis of a hydrogen-fueled turbofan engine. In this article, an economical optimization and environmental optimization of the turbofan engine with hydrogen fuel was performed. The results of this study indicate that for the optimized cycle with hydrogen fuel usage, the thermal efficiency increase by 2.65%. Balli et al.[11] examined the efficacy of hydrogen fuel usage on the exergetic performance of a turbojet engine. The results of this study show that using hydrogen fuel exergetic efficiency of the engine from 15.40 to 14.33 percentage decreases. In another study, Gaspar and Souca[12] examined the effect of environmental performance of a small turbofan engine. Optimization is done to find the appropriate values for the design variable of the cycle in order to improve performance parameters and environmental performance. Dik et al.[13] designed a conceptual design and optimization of a three spools turbofan engine with the function of reducing fuel consumption for operation in 2025. Compared to the base engine, their results show this the fan diameter and the overall length of the engine increased by 21 and 2.2 percent, respectively, and its fuel consumption rate decreased by 11 percent. Li and Tan[14] performed thermodynamic analysis to find the optimum compression ratio in the split of intercooled recuperated turbofan. They found that the split of intercooling system pressure ratio affects the compression ability of compressors. In another study, Yucer[15] performed the thermodynamic analysis of the cycle of a small jet engine based on the exergy. They also examined the performance of the engine load part. In their study, second law efficiency and exergy destruction rates of different engine subsystems were calculated. The results of this paper showed that the highest exergy destruction rate occurs in the combustor, and TSFC occurs in the full load condition. Balli[16] modeled the exergy modeling of a high by-pass ratio turbofan engine used in business airplains. The resistance of the PW4056 engine in Different operating conditions was investigated. The results of this study indicate that the engine operates in MTOP operating mode with a better level of stability. Zhao et al.[17] investigated the second and first lows of energy analysis for the intercooled turbofan engine in different flight states. The results showed the highest exergy destruction rate belongs to the combustor among engine components. Also, the intercooling method reduces the exergy destruction rate of the combustor. Tuzcu et al.[18] conducted an economic and environmental analysis of a gas turbine engine used in the aerospace industry. The results of this study show that the energy efficiency of combustor, low-pressure turbine (LPT), and high-pressure turbine (HPT) are 90, 26.2, and 19.7%. respectively. Also, CO2 emission per day is calculated at 358.9 tons per day. In another study, S.D. Farahani et al.[19] optimized the performance of the TF30-P414 low by-pass turbofan engine. The results of this study indicate that the optimum second law efficiency of the engine is 32.64 in flight condition included Flight-Mach number of 1.944 and Flight altitude of 11.236 meters. In another study, M.Sabzehali et al.[20] developed a model via deep learning to predict the energy and exergy performance of the F135 PW100 turbofan engine with the aim of optimizing the energy and exergy performance for flying at an altitude of 30,000 meters and Mach 2.5.

In another study, Ekrataleshian et al.[21] performed exothermic thermo-economic analysis and exergetic optimization of a turbojet engine. Final decision-making methods of Technique for Order of Preference by Similarity to Ideal Solution (TOPSIS) and linear programming technique for multidimensional analysis of preference (LINMAP) were used to select the optimal cycle. The energy efficiency of the chosen optimal cycle with TOPSIS is 65.86%. Xu et al.[22] analyzed a turbofan engine with a new re-cooled system. The results show that the novel re-cooled increases the thrust force, especially in the upper airways.

In previous studies, Farahani[19] et al. optimized the performance of TF30-P414 turbofan engine with low bypass ratio using ( teaching-learning process inspired) TLBO algorithm and based on the analysis of energy and exergy of the engine. The difference between this study and the current study is the use of the TLBO algorithm for optimization and different design variables and their change ranges for optimization and the studied engine. Also in another study, M.Sabzehali[20] et al. developed a model based on deep learning to predict the energy and exergy performance of the F135 PW100 mixed flow turbofan engine in the conditions of altitude of 30000 meters and Mach 2.5. Among the differences of this study[20] compared to the current study is the different engine that the F135 Pw100 is a low bypass mixed flow turbofan, However, the current engine is GENX 1B70 engine that is a high bypass un-mixed flow turbofan engine, Also, the working conditions, methods and algorithms studied and the purpose of the study are different. In this study, it is only a prediction model based on deep learning to predict the performance of the F135 PW100 turbofan engine, which is a mixed flow turbofan engine with a low bypass ratio was developed, but in the present study, the aim is to optimize the performance of GENX 1B70 un-mixed flow turbofan engine with high bypass ratio.

In this study, the GENX 1B70 engine was modeled using MATLAB software. The efficacy of inlet air cooling and fuel type on the GENX 1B70 turbofan engine performance in two take-off conditions and on design conditions is studied. The design cycle of the GENX 1B70 engine cycle with separate target functions of thrust maximization, thermal efficiency maximization, and propulsive efficiency maximization based on the Genetic algorithm was investigated. Then based on the TOPSIS algorithm and using MATLAB software were selected from the optimized cycles of the best cycles with two different approaches.

## 2 .Materials and Methods

A separated flow TF engine with an inlet-air cooling system schemTatic is illustrated in Figure 1. Engine components include fan, HPC, LPC, combustor, HPT, LPT, and cooling system. The inlet airflow temperature decreases after passing through the heat exchanger of the inlet-air cooling system. Then the cooled air arrives the low-pressure compressor and fan. The output flow of the low-pressure compressor enters the high-pressure compressor. Then the output flow of the high-pressure compressor enters the combustion chamber. The combustion products enter the high-pressure turbine and then enter the low-pressure turbine. The required power of the low-pressure compressor and fan is supplied by the low-pressure turbine and the required power of the high-pressure compressor is supplied by the high-pressure turbine. The low-pressure turbine exit flow enters the hot stream nozzle and then it exhausts to the ambient. And output flow of the fan enters the cold stream nozzle, then it exhausts to the ambient.[23]

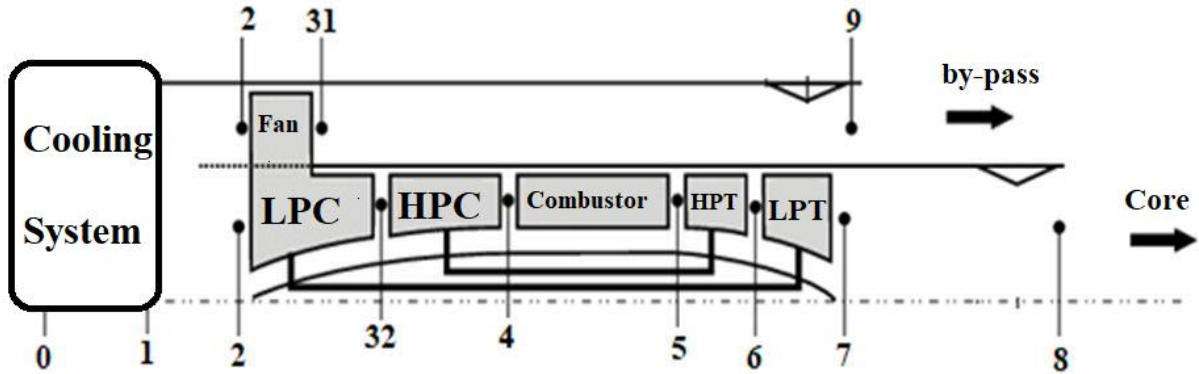

**Figure 1.** A schematic of a separated flow TF engine with an inlet air cooling system model configuration.[23]

**2.1. Energy Analysis**

To add a cooling system to the engine system, it is recommended to use the engine intake duct as a cooling converter. The airflow pressure changes at the cooling system and intake are ignored[20].

$$P_0 = P_1 \tag{1}$$

$$T_1 = T_0 + \Delta T \tag{2}$$

$P_0$, $T_0$, $P_1$, and $T_1$ are ambient air pressure, ambient air temperature, diffuser input pressure, and diffuser input temperature, respectively. $P_2$ and $T_2$ are the fan input pressure and fan input temperature respectively, and $\Delta T$ is $T_1-T_0$ based on International Standard Atmosphere (ISA) [24].

Diffuser output air temperature is calculated as follows[1].

$$T_2 = T_1 \left(1 + \frac{k_{diff} - 1}{2} Ma^2\right) \tag{3}$$

The diffuser output flow pressure is premeditated as follows[1].

$$P_2 = P_1 \left(\frac{T_2}{T_1}\right)^{\frac{k_{diff}}{k_{diff}-1}} \tag{4}$$

Fan output flow pressure is obtainable from the following equation[1].

$$P_{31=\pi_{fan}} P_2 \tag{5}$$

where $\pi_{fan}$ is fan pressure ratio and $P_{31}$ is fan output flow pressure and $P_2$ is fan input flow pressure, also, fan output flow temperature is premeditated as follows[1]:

$$T_{31} = \frac{T_2}{\eta_{fan}}\left[\left(\frac{P_{31}}{P_2}\right)^{\frac{k_F-1}{k_F}} - 1\right] + T_2 \tag{6}$$

where $T_{31}$ is fan output air temperature, the output flow pressure of the low pressure compressor is obtained as follows[1]:

$$P_{32} = \pi_{Cl} P_2 \tag{7}$$

where $\pi_{Cl}$ is the low-pressure compressor pressure ratio and $P_{32}$ is the exit pressure of the low-pressure compressor. The exit temperature of the low pressure compressor is obtained as following[1]:

$$T_{32} = \frac{T_2}{\eta_{Cl}}\left[\left(\frac{P_{32}}{P_2}\right)^{\frac{k_{cl}-1}{k_{cl}}} - 1\right] + T_{31} \tag{8}$$

where $T_{32}$ is low pressure compressor exit air temperature and $T_2$ is low pressure compressor input flow temperature. The output flow pressure of the high pressure compressor is obtained as follows[1]:

$$P_4 = \pi_{Ch} P_{32} \tag{9}$$

where $\pi_{Ch}$ is high pressure compressor pressure ratio and $P_4$ is the exit pressure of the high-pressure compressor. The exit temperature of the high pressure compressor is obtained as following[1]:

$$T_4 = \frac{T_{32}}{\eta_{Ch}}\left[\left(\frac{P_4}{P_{32}}\right)^{\frac{k_{ch}-1}{k_{ch}}} - 1\right] + T_{32} \tag{10}$$

where $T_4$ is high pressure compressor exit air temperature and $T_{31}$ is fan exit air temperature. Intake mass flow rate is premeditated as follows:

$$m_t = \rho V_0 A \tag{11}$$

where A is a aera of the intake flow, and $\rho$ is the density of intake flow and $V_0$ is flight velocity. The low pressure compressor input mass flow rate is obtained as follows[1]:

$$m_h = \frac{m_t}{\alpha + 1} \tag{12}$$

where $m_h$ is the low pressure compressor compressors input mass flow rate. High pressure compressor mechanical power is attainable as follows[1]:

$$W_{Ch} = m_h \frac{T_{32}}{\eta_{Ch}}\left[\left(\frac{P_4}{P_{32}}\right)^{\frac{k_{ch}-1}{k_{ch}}} - 1\right] \tag{13}$$

where $W_{Ch}$ is High pressure compressor power consumption.

The mechanical power of low pressure compressor is obtained as follows[1]:

$$W_{Cl} = m_h \frac{T_2}{\eta_{Cl}} \left[ \left(\frac{P_{32}}{P_2}\right)^{\frac{k_{cl}-1}{k_{cl}}} - 1 \right] \tag{14}$$

where $W_{Cl}$ is low pressure compressor mechanical power consumption.

Fan mechanical power consumption is gotten as follows[1]:

$$W_f = m_t \frac{T_2}{\eta_{fan}} \left[ \left(\frac{P_{31}}{P_2}\right)^{\frac{k_f-1}{k_f}} - 1 \right] \tag{15}$$

where $W_f$ is mechanical power consumption of the fan and $m_t$ is input mass flow rate of the fan. By-pass mass flow rate is obtained as follows[1]:

$$m_c = m_t - m_{ah} \tag{16}$$

where $m_c$ is by-pass actual mass flow. combustor exit flow pressure is obtained as follows[1]:

$$P_5 = P_4 - \Delta P_{combustor} \tag{17}$$

where $P_4, P_5$, and $\Delta P_{combustor}$ are high pressure compressor exit pressure, combustion chamber exit pressure, and pressure drop in the combustion chamber, respectively. Heat rate is calculated as follows[1]:

$$HR = m_h C_{avcc}(T_5 - T_4) \tag{18}$$

where $HR$ is the heat production rate, $T_4$ is the input flow temperature of the combustor, and $T_5$ is the combustor exit flow temperature. Fuel mass flow rate is calculated as following equation[1]:

$$m_f = \frac{HR}{FHV \, \eta_{combustor}} \tag{19}$$

where FHV is specific fuel heat value, and $\eta_{combustor}$ is combustion efficiency at the combustor, and $m_f$ is fuel mass flow rate, and $HR$ is the heat production rate. HPT inlet mass flow rate is computed as following[1]:

$$m_{Turb} = m_h + m_f \tag{20}$$

where $m_f$ is the fuel consumption rate.

HPT mechanical power is computed as follows[1]:

$$W_{HPT} = m_{Turb} C_{Pavhpt}(T_5 - T_6) \tag{21}$$

where $m_{Turb}$ is the HPT inlet mass flow rate. low- pressure turbine (LPT) mechanical power is calculated as following equation[1]:

$$W_{LPT} = m_{Turb} C_{Pavlpt} \cdot (T_6 - T_7) \tag{22}$$

where $m_{Turb}$ is the LPT inlet mass flow. According to the law of energy conservation, the HPT mechanical power is equal to mechanical power consumption of the high pressure compressor:

$$m_{Turb}C_{Pavhpt}\cdot(T_5 - T_6) = m_h\left[\frac{T_{32}}{\eta_{Ch}}\left[\left(\frac{P_4}{P_{32}}\right)^{\frac{k_{ch}-1}{k_{ch}}} - 1\right]\right] \quad (23)$$

where $T_6$ and $T_5$ are the HPT output and input flow temperature respectively, $m_{Turb}$ is the HPT input mass flow rate. The output flow temperature of the HPT is obtainable at equation (21), the HPT output flow pressure is calculated as follows[1].

$$P_6 = P_5\left[1 - \frac{1}{\eta_{HPT}}\left(1 - \frac{T_6}{T_5}\right)\right]^{\frac{k_{hpt}}{k_{hpt}-1}} \quad (24)$$

where $T_6$ is the output flow temperature of the HPT, $T_5$ is the input flow temperature of the HPT, and $P_6$ and $P_5$ respectively, are the high pressure turbine output flow pressure and high pressure turbine input flow pressure. $k_{hpt}$ is the ratio of specific heat at constant pressure to specific heat at constant volume and HPT isentropic efficiency is $\eta_{HPT}$. According to the law of energy conservation, the LPT mechanical power is equal to mechanical power consumption of the fan and low pressure compressor:

$$m_{Turb}C_{Pavlpt}(T_6 - T_7) = m_t\left[\frac{T_2}{\eta_{fan}}\left[\left(\frac{P_{31}}{P_2}\right)^{\frac{k_f-1}{k_f}} - 1\right]\right] + m_h\frac{T_2}{\eta_{Cl}}\left[\left(\frac{P_{32}}{P_2}\right)^{\frac{k_{cl}-1}{k_{cl}}} - 1\right] \quad (25)$$

where $T_7$ is the LPT output temperature, and LPT input temperature is the $T_6$, $m_T$ is the fan inlet mass flow rate. by obtaining the LPT exit temperature of the equation (22), The flow pressure at the LPT exit is obtained[1]:

$$P_7 = P_6\left[1 - \frac{1}{\eta_{LPT}}\left(1 - \left(\frac{T_8}{T_7}\right)\right)\right]^{\frac{K_{nh}}{K_{nh}-1}} \quad (26)$$

where $P_7$ is the flow pressure at the LPT output and $T_7$ is the flow temperature at the LPT output. The hot stream nozzle exit pressure is also calculated in such a way[1]:

$$P_8 = P_7\left[1 - \frac{1}{\eta_{nh}}\left(1 - \left(\frac{T_8}{T_7}\right)\right)\right]^{\frac{K_{nh}}{K_{nh}-1}} \quad (27)$$

where $T_8$ is the hot stream nozzle output flow temperature, and $P_7$ is the hot stream nozzle input flow pressure. The cold stream nozzle exit pressure is also calculated in such a way[1]:

$$P_9 = P_{31}\left[1 - \frac{1}{\eta_{nc}}\left(1 - \left(\frac{T_9}{T_{31}}\right)\right)\right]^{\frac{K_{nc}}{K_{nc}-1}} \quad (28)$$

where $T_9$ is the cold stream nozzle exit flow temperature, and $P_9$ is the cold stream nozzle exit flow pressure. The hot stream nozzle exit velocity is obtained in such a way[1]:

$$V_8 = \left(2\eta_{nh}\frac{K_{nh}}{K_{nh}-1}RT_7\left[1 - \left(\frac{P_8}{P_7}\right)^{\frac{K_{nh}-1}{K_{nh}}}\right]\right)^{0.5} \quad (29)$$

where R is the gas global constant, and $P_8$ is the hot stream nozzle exit pressure. The cold stream nozzle exit velocity is calculated in such a way[1]:

$$V_9 = \left(2\eta_{nc}\frac{K_{nc}}{K_{nc}-1}\cdot R.T_{31}\left[1-\left(\frac{P_9}{P_{31}}\right)^{\frac{K_{nc}-1}{K_{nc}}}\right]\right)^{0.5} \tag{30}$$

where $P_9$ is cold stream nozzle exit pressure. Flight velocity is calculated as:

$$V_0 = Ma(RkT_0)^{0.5} \tag{31}$$

The hot stream nozzle thrust is calculated in such a way[1]:

$$F_{hot} = \frac{1}{g_c}\left[\left((m_{ah}+m_f)(V_8)\right)-(m_{ah}V_0)\right]+[A_8(P_8-P_0)] \tag{32}$$

where $A_8$ is the hot stream nozzle exit area, and $g_c$ is the gravity of the earth and $F_{hot}$ is the hot stream nozzle thrust force.

The cold stream nozzle thrust is calculated as[1]:

$$F_{cold} = \frac{1}{g_c}\left[\left((m_c)(V_9)\right)-(m_cV_0)\right]+[A_9(P_9-P_0)] \tag{33}$$

where $V_9$ is the cold stream nozzle exit area, $P_9$ is the pressure at the cold nozzle output, and $F_{cold}$ is the cold stream nozzle thrust force.

The total thrust is computed.

$$F_{total} = F_{hot} + F_{cold} \tag{34}$$

where $F_{total}$ is total thrust. thrust specific fuel consumption (TSFC) is the ratio of fuel mass flow rate to total thrust[1]:

$$TSFC = \frac{m_f}{F_{total}} \tag{35}$$

Thermal efficiency is the ratio of the flow kinetic energy in the engine to the total heat added to the flow during the combustion chamber of the engine[1].

$$\eta_{th} = \frac{\left[(m_h+m_f)V_8^2-(m_tV_0^2)+m_c(V_9^2-V_0^2)\right]}{2m_fFLV} \tag{36}$$

where $\eta_{th}$ is the thermal efficiency. Propulsive efficiency is the ratio of the total thrust at the flight speed to the flow kinetic energy changes in the engine[1].

$$\eta_p = \frac{F_{total}V_0}{\left[(m_h+m_f)V_8^2-(m_tV_0^2)+m_c(V_9^2-V_0^2)\right]} \tag{37}$$

where $\eta_p$ is the propulsion efficiency. The overall efficiency ($\eta_o$) is calculated as[1]:

$$\eta_o = \eta_p \eta_{th} \tag{38}$$

TSF is calculated in such a way[1]:

$$TSF = \frac{F_{total}}{m_t} \tag{39}$$

where $F_{total}$ and $m_t$ are total thrust, and intake real air mass flow rate. The nitrate oxide emission index coefficient (SNOx) is obtained in such a way[25, 26]:

$$S_{NOx} = \left(\frac{P_4}{2965[kPa]}\right)^{0.4} e^{\left(\frac{T_4-826[K]}{194[K]} + \frac{6.29-(100.war)}{53.2}\right)} \tag{40}$$

where *war* is the ratio of the liquid water- air to the compressor exit air. In this study, the effect of the combustor geometry and fuel type on SNOx has been omitted, and only the effect of temperature, pressure and liquid water-air ratio of the incoming air on it has been discussed. The mass flow rate of nitrogen oxide produced by the engine is calculated as a coefficient of Snox, and the fuel mass flow rate[25].

$$m_{NOx} = 23 S_{NOx} m_f \tag{41}$$

where $m_f$ is the mass flow rate of the engine consumed, and $m_{NOx}$ is the mass flow rate of the nitrogen oxides produced by the engine and its unit is in grams per second.

**2-2. Exergy Analysis modeling**

Exergy is the maximum beneficial work that can be achieved from the system. The physical exergy for airflow at the inlet and outlet of all engine components is calculated as follows[27, 28]:

$$\psi = (S - S_0) - T_0(H - H_0) \tag{42}$$

where S and $S_0$ are the fluid flow specific entropy and the environment specific entropy, respectively, and H and $H_0$ are the fluid flow specific enthalpy and the environment specific enthalpy, respectively. Also, $T_0$ is the temperature of ambient air. The airflow physical exergy rate is calculated as follows[27]:

$$\Psi_a = m_a \psi \tag{43}$$

where $\psi$ is the specific physical exergy of the air flow, the chemical exergy rate of fuel flow is calculated as follows[29]:

$$\Psi_f = m_f \Psi_x \tag{44}$$

where $\psi_x$ is the specific chemical exergy of the fuel flow, and $m_f$ is the fuel consumed mass flow rate and $\Psi_f$ is the rate of fuel flow chemical exergy. Hydrocarbon fuel Specific chemical exergy is calculated as follows[29]:

$$\psi_x = FHV\left(1.04224 + \frac{0.11925c}{h} - \frac{0.042}{c}\right) \tag{45}$$

Where c, h and FHV are the number of carbon and hydrogen atoms in each molecule of the fuel, and the fuel heat value, respectively. The exergetic efficiency of the aero gas turbine is calculated as follows[29]:

$$\eta_{ex} = \frac{F_{total} v_o}{\Psi_f} \tag{46}$$

where $\eta_{ex}$ is the exergetic efficiency of the gas turbine engine, the exergy efficiency of the fan is calculated as follows[29]:

$$\eta_{exf} = \frac{\Psi_{Of} - \Psi_{if}}{W_f} \tag{47}$$

where $\Psi_{Of}$, and $\Psi_{if}$ are the fan exit and inlet exergy rate, respectively and $W_f$ is mechanical power consumption of the fan. Exergy destruction rate of the fan ($E_{Df}$) is calculated in this way[29]:

$$E_{Df} = W_f + \Psi_{if} - \Psi_{Of} \tag{48}$$

The compressor exergetic efficiency is calculated as follows[29]:

$$\eta_{exc} = \frac{\Psi_{oc} - \Psi_{ic}}{W_c} \tag{49}$$

where $\Psi_{oc}$, $\Psi_{ic}$ and $W_c$ are the compressor output flow exergy rate, compressor input flow exergy rate, and the mechanical power consumption of the compressor, respectively. Also, the rate of exergy destruction in the compressor ($E_{DC}$) is computed as follows[29]:

$$E_{DC} = W_c + \Psi_{ic} - \Psi_{oc} \tag{50}$$

The exergy efficiency of the turbine is computed in this way[29]:

$$\eta_{exT} = \frac{W_T}{\Psi_{it} - \Psi_{ot}} \tag{51}$$

where $W_T$ is the turbine power output, and $\eta_{exT}$ is the turbine exergy efficiency. where $\Psi_{it}$ and $\Psi_{ot}$ are the turbine input, and output flow exergy rate, respectively. Also, the rate of exergy destruction ($E_{DT}$) in the turbine is calculated in this way[29]:

$$E_{DT} = \Psi_{it} - \Psi_{ot} - W_T \tag{52}$$

The combustor exergy efficiency is also calculated in this way[29]:

$$\eta_{exCC} = \frac{\Psi_{occ}}{\Psi_{icc} - \Psi_f} \tag{53}$$

where $\Psi_f$ is the chemical exergy rate of the fuel flow and $\eta_{exCC}$ is the combustor exergy efficiency. Also, $\Psi_{icc}$ and $\Psi_{occ}$ are combustor input and output flow exergy rate, respectively. The rate of exergy destruction in the combustor ($E_{Dcc}$) is calculated in this way[29]:

$$E_{Dcc} = \Psi_{icc} - \Psi_{occ} + \Psi_f \tag{54}$$

## 2-3. Validation

The components of GENX 1B70 engine include one stage axial flow fan, four stages low- pressure axial flow compressor, ten stages high- pressure axial flow compressor, two stages axial flow high- pressure turbine, and seven stages axial flow low pressure turbine. This engine has been used as a propulsion system for aircraft such as Boeing 747 and Boeing787-8[30]. The input parameters for GENX 1B70 engine are given in Table 1[30].

**Table 1. The input parameters for GENX 1B70 engine.**

| Parameters | Value | Unit |
|---|---|---|
| TIT | 1695 | K |
| $\pi_{fan}$ | 1.5 | - |
| $\pi_{LPC}$ | 1.3 | - |
| HP compressor pressure ratio | 23 | - |
| $\alpha$ | 9.1 | - |
| $m_a$ | 1155.43 | Kg/s |
| $\eta_{fan}$ | 0.91 | - |
| LP compressor isentropic efficiency | 0.91 | - |
| HP compressor isentropic efficiency | 0.91 | - |

The efficacy of hydrogen, natural gas, and JP10 has also been investigated on the performance of the engines considered. The thermal value (FHV) and chemical exergy per mass unit are given for each fuel in Table 2[31].

**Table 2. FHV and chemical exergy per mass unit for each fuels.**

| Fuel type | JP10 | NATURAL GAS | HYDROGEN |
|---|---|---|---|
| chemical exergy per mass unit (MJ/Kg) | 44.921 | 55.168 | 134.778 |
| FHV (MJ/Kg) [31] | 42.075 | 49.736 | 118.429 |
| Molecular weight(g/mol) [31] | 136 | 16 | 2 |

|                      |            | $C_{10}H_{16}$ | $CH_4$ | $H_2$ |
|---|---|---|---|---|
| Chemical formula [31] | | | | |

In this paper efficacy of inlet air cooling on GENX 1B70 engine in take-off condition (Ma = 0, H = 0) has been investigated. First, modeling validation is discussed and the results obtained for thrust force and TSFC with the values announced in the references for GENX 1B70, engine in take-off (Ma = 0, H = 0) and on design (Ma = 0.85, H = 10000 m) conditions for JP10 fuel is compared in Table 3. According to the calculated error percentage, it can be seen that the results of modeling for the engine are in acceptable agreement with references, and less than 10% of the error is observed.

**Table 3. Comparison of the results of the present study with reference.**

|  | Parameters | Take off condition | On design condition |
|---|---|---|---|
| **Thrust (KN)** | Present study | 310 | 72.5 |
|  | Error (percent) | 2.25 | 5.92 |
|  | Reference value [30] | 317 | 76 |
| **TSFC(g/KNs)** | Present study | 8.454 | 18.001 |
|  | Error (percent) | 8.60 | 7.78 |
|  | Reference value [30] | 9.25 | 19.52 |

## 2.4. Optimization Methodology

Genetic algorithm is an innovative and optimized search method inspired by Charles Darwin's theory of natural selection. This algorithm actually represents the theory of natural selection; Where the most suitable people are selected to continue the generation and produce children. The function of choosing nature begins with selecting the best people from a population. These people produce offspring who inherit their parental characteristics and pass them on to the next generations. If parents adopt better, their children will be better and more adaptable than their parents and will have an even better chance of survival. This process will be repeated continuously and eventually a generation will be created that will be most compatible with the environment. Nature is always a very important source of ideas and inspiration to solve many of our problems. Genetic algorithm (GA) is one of the same ideas that humans have extracted from the heart of nature. In this method, a set of solutions and ideas to solve problems is considered, ideas and solutions are combined with each other and in some cases mutate and produce new children (new solutions) and this process during different generations. It is constantly repeated and each point or solution is assigned a point. In the end, better and more efficient solutions that get higher scores have a better chance of being chosen by nature. There are five phases for the genetic algorithm: initial population, Fitness function, selection, crossover, and mutation. The algorithm terminates when the population converges to a specific instance and solution. That is, in fact, children are not eventually different from their parents. In this case, it can be said that the genetic algorithm has provided a set of solutions to our problem. This sequence of phases and phases is repeated to create people who are better and more adaptable than their predecessors. In this study, optimization of the GENX 1B70 turbofan engine cycle with three separate objective functions as a single objective function based on the genetic algorithm was performed using MATLAB software. The target functions are thrust maximization, thermal efficiency maximization, and propulsive efficiency maximization. The flowchart of methodology and GA is provided in Figure 2. In this study, optimization of the GENX 1B70 turbofan engine cycle with three separate objective functions as a single objective function based on the genetic algorithm was performed using MATLAB software. The target functions are thrust maximization, thermal efficiency maximization, and propulsive efficiency maximization[32].

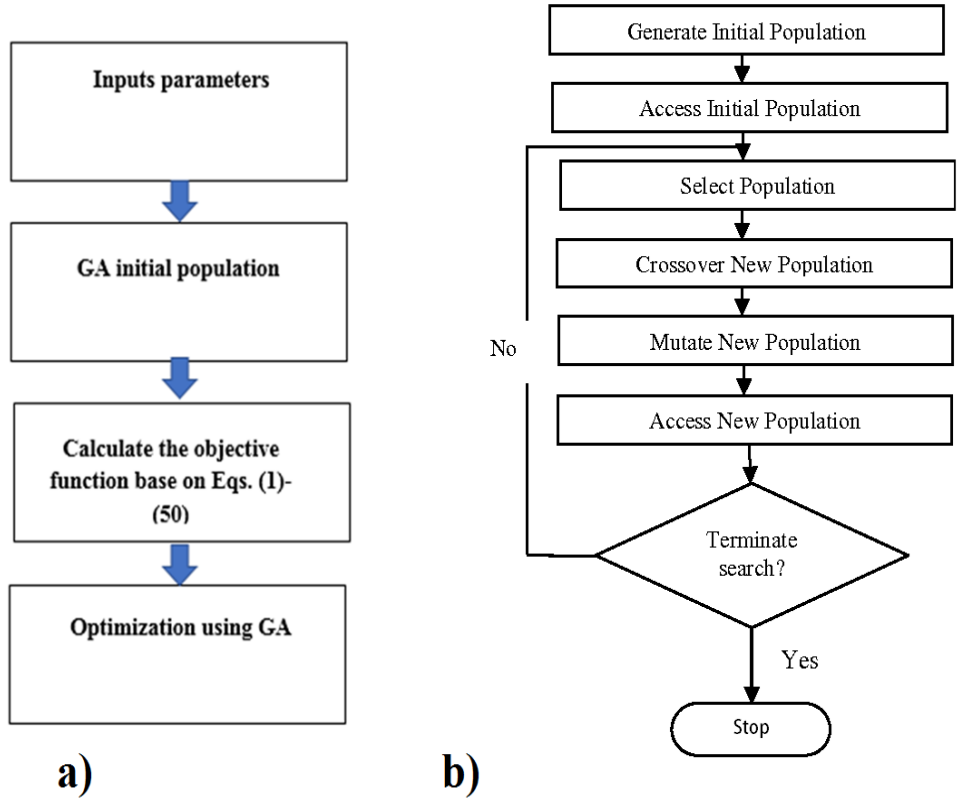

**Figure 2.** a) flowchart of methodology and b) Genetic algorithm.

## 2.5. TOPSIS Method

In this study, among the optimized cycles based on GENX B70 engine cycle with approach, Economic and Exero- Environmental, by determining the appropriate weight coefficients based on Technique for Order of Preference by Similarity to Ideal Solution algorithm and using MATLAB software has been done. In optimization methods based on several objective functions, instead of one answer, a series of candidate answers are obtained for the final answer, which should be selected using the appropriate strategy. In this case, the conventional TOPSIS method with Euclidean dimensioning has been used[32]. This model is one of the best multi-character decision models. The concept used in this method is that the selected option should have the shortest distance with the best possible mode and the longest distance with the worst-case scenario. Problem-solving with this method involves six steps[33].

1- Forming a problem decision matrix.

2- Forming a scale-less matrix. The decision-making matrix is converted to a non-scale matrix using the following Euclidean norm.

$$n_{ij} = \frac{r_{ij}}{\left(\sum_{i=1}^{m} r_{ij}^2\right)^{1/2}}, (i = 1,2,\ldots,m), (j = 1,2,\ldots,n) \tag{55}$$

$$N_d = [n_{ij}] \tag{56}$$

3- Forming a rhythmic scale less matrix: In this step, the weight of each of the previously calculated indicators is multiplied in each of the options, and the weighted scale less scale is calculated as follows:

$$V = N_d \times W_{n \times n} \tag{57}$$

In the above equation, V is an unbalanced scale less matrix, and W is a diagonal matrix of the weights obtained for the indicators.

4- Calculating the distance (d) or proximity to the answer of the positive ideal $(^+A_i)$ and the negative ideal $(^-A_i)$ based on the Euclidean norm, for the positive indicators $J_1 = \{1,2,..,n\}$ and the negative indicators $J_2 = \{1,2,..,n\}$.

$$d_i^+ = \left\{\sum_{j=1}^{n}\left(V_{ij} - V_j^+\right)^2\right\}^{\frac{1}{2}}, (i = 1,2,\ldots,m)$$

$$d_i^- = \left\{\sum_{j=1}^{n}\left(V_{ij} - V_j^-\right)^2\right\}^{\frac{1}{2}}, (i = 1,2,\ldots,m) \tag{58}$$

5- Determining the relative proximity: The relative proximity of the option to the ideal solution is calculated as follows[33]:

$$C_i = \frac{d_i^-}{d_i^- + d_i^+}, (i = 1.2,\ldots,n) \tag{59}$$

6- Ranking of options: Based on the relative values of the options in this stage, based on the descending order, the available options can be ranked based on the most important.

### 3-Results and Discussion

In this paper, first, the GENX 1B70 engine was modeled with MATLAB software. Then, the results obtained from this modeling in two take-off conditions, and on design conditions were compared with reference values. The efficacy of inlet air temperature on GENX 1B70 engine performance and exergetic efficiency of components of GENX 1B70 turbofan engine in two conditions of take-off (Ma = 0, H = 0) and on design conditions (Ma = 0.85, H = 10000 m) was studied. Also, the efficacy of fuel type on the GENX 1B70 engine performance and the exergetic efficiency of the GENX 1B70 engine on the design point condition was investigated.

#### 3.1. Energy Analysis

The engine input mass flow rate variations with input air flow temperature variations in both take-off, and on design conditions are shown in Figure 3. Lessening the input air flow temperature increases the input air density and it increases the rate of input air mass flow as the air density increases. Increasing the flow of air entering the engine increases the thrust force. It has also been observed; the engine has the highest thrust force with the use of hydrogen fuel that is also observed by M.Sabzehali et al.[20]

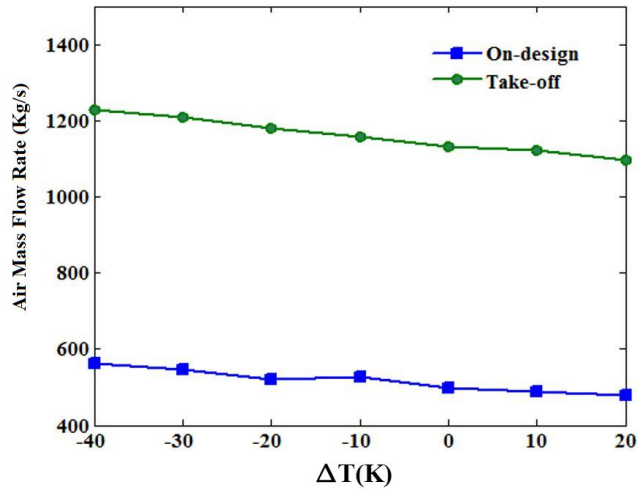

**Figure 3.** The engine input mass flow rate variations with input air flow temperature changes in both take- off and on design conditions

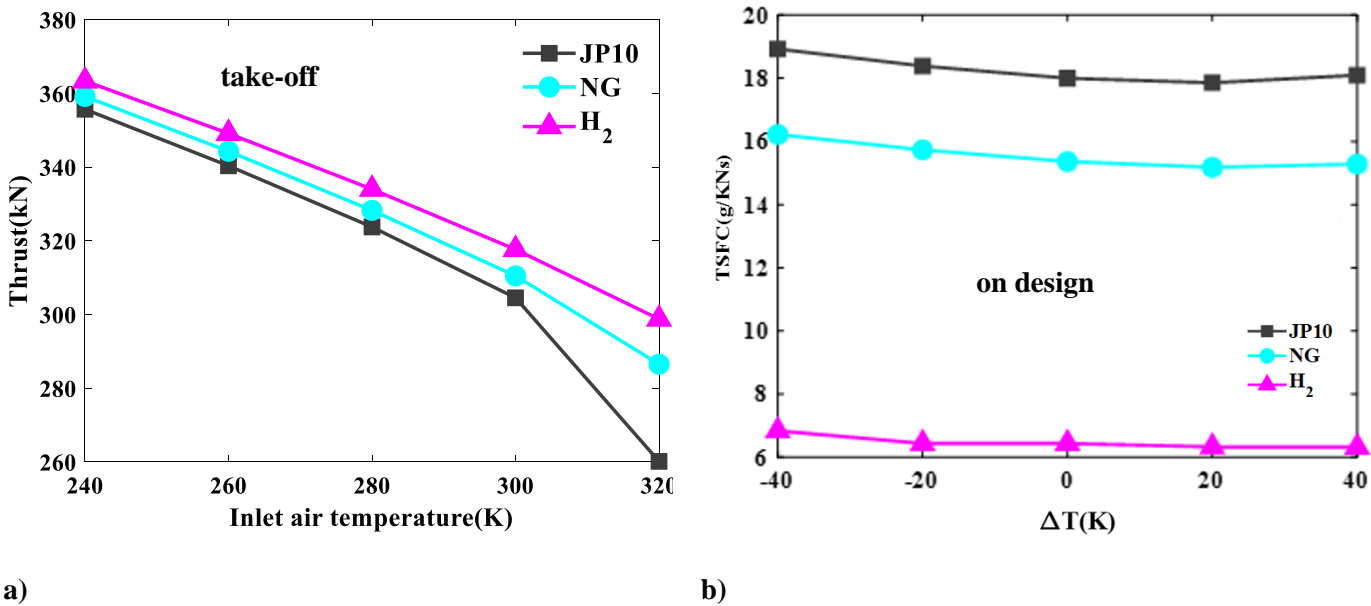

a)  b)

**Figure 4.** Thrust force changes with changes in inlet air flow temperature with a) take-off and b) on design conditions for GENX 1B70 engine

Thrust force changes with changes in inlet air temperature with take-off and on design conditions for GENX 1B70 engine is shown in Figure 4.The engine fuel flow mass rate changes with inlet air temperature in the take-off (Ma=0, H=0 ), and on design (Ma=0.85, H=10000 m) conditions for the GENX 1B70 engine are shown in Figure 5. As a result, the drop of inlet air temperature under equal conditions, higher heat rate is required to bring the temperature of each kilogram of inlet air to the turbine inlet temperature. With a drop in inlet air temperature, the inlet air mass flow rate also increases; as a result, the fuel mass flow rate increases. It has also been observed that as fuel heat value (FHV) increases, fuel mass flow rate decreases because the heat rate is constant under equal conditions. Therefore, the thermodynamic cycle for hydrogen fuel has the lowest fuel mass flow rate compared to other fuels.

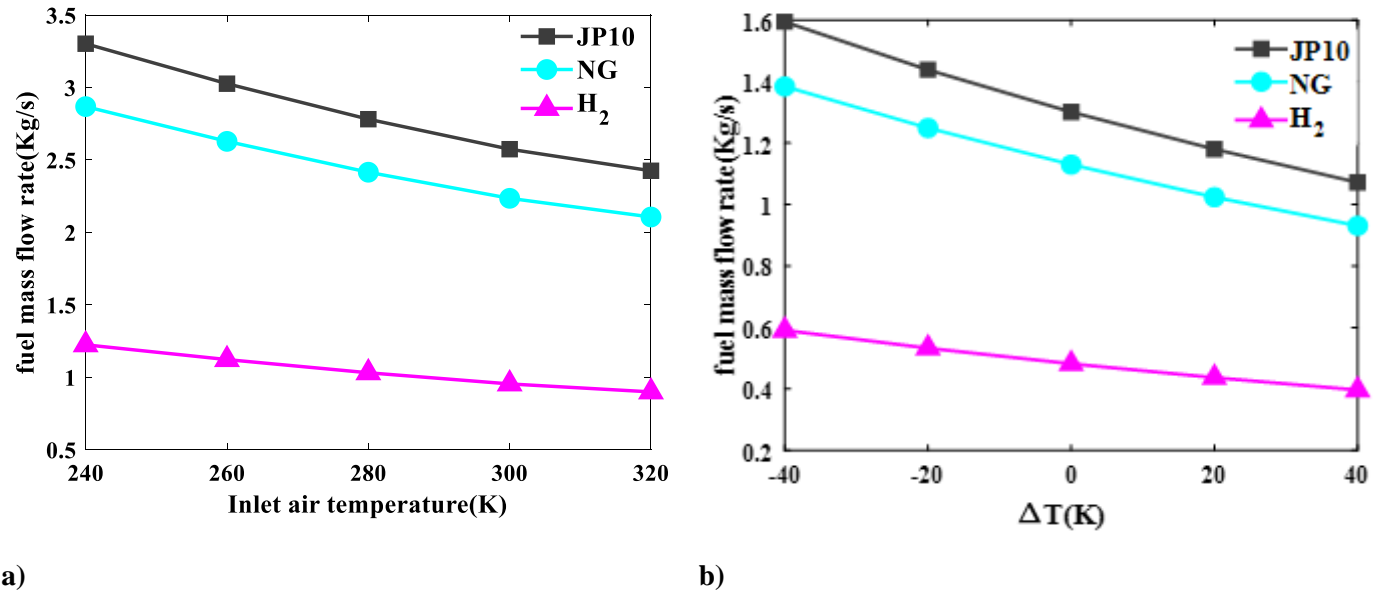

a)  b)

**Figure 5.** The engine fuel flow mass rate changes with inlet air flow temperature in a) the take-off and b) on design conditions for the GENX 1B70 engine

Figure 6 shows thermal efficiency changes with inlet air temperature changes with take-off condition (Ma = 0, H = 0) and on design condition (Ma=0.85, H=10000 m) for GENX 1B70 engine. As the inlet air temperature decreases, the rate of inlet air mass flow increases. Therefore, the rate of flow kinetic energy changes along with the engine increases. Where the rate of grow in the kinetic energy changes, the rate of airflow of the engine due to the decrease in inlet air temperature is greater than the rate of grow in heat rate due to the decrease in inlet air temperature. Also, thermal efficiency increases with reducing inlet flow temperature.

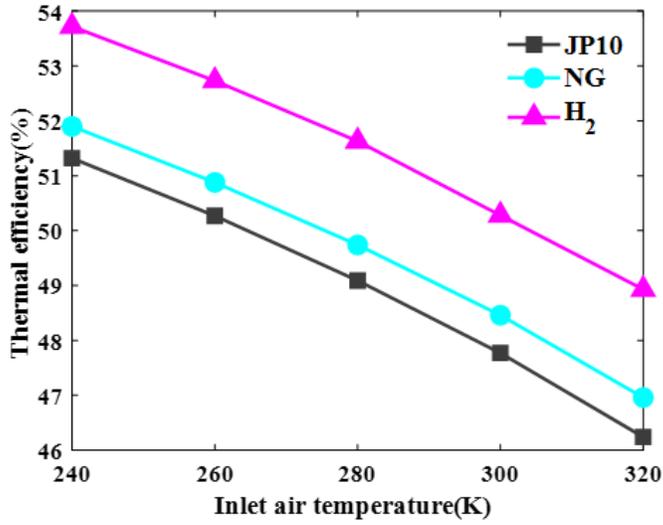 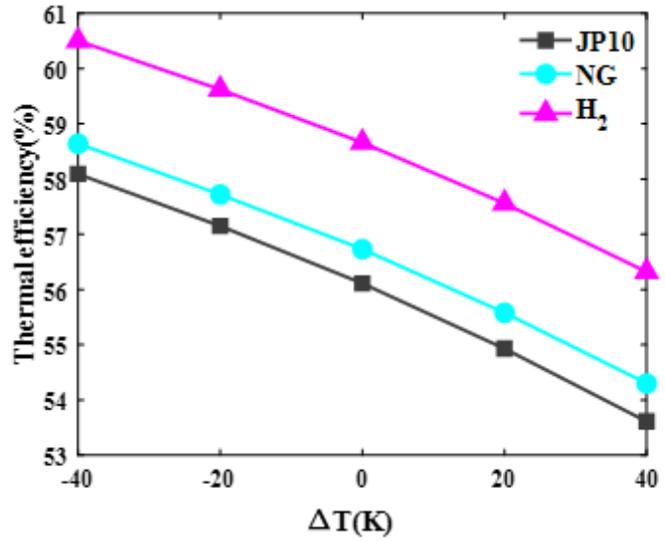

a)  b)

**Figure 6.** thermal efficiency changes with inlet air flow temperature changes with a) take-off condition and b) on design condition for GENX 1B70 engine

Figure 7 shows TSFC for inlet air mass flow rate changes in take-off condition and on design condition for GENX 1B70, engine. It is observed that in the take-off condition in order to reduce the inlet air temperature in the temperature range of less than 290 K, the thrust force changes rate is less than the rate of grow of the fuel mass flow rate. Therefore, the TSFC increases. It is observed that in the on-design condition, in order to reduce the inlet air temperature in the temperature range of less than 253 K (i.e., Δ T=20 K), the thrust force changes rate is less than the rate of increase of the fuel mass flow rate. Therefore, the TSFC increases. Since the heat rate is stable under equal conditions, TSFC decreases with increasing fuel heat value, so hydrogen has the lowest TSFC.

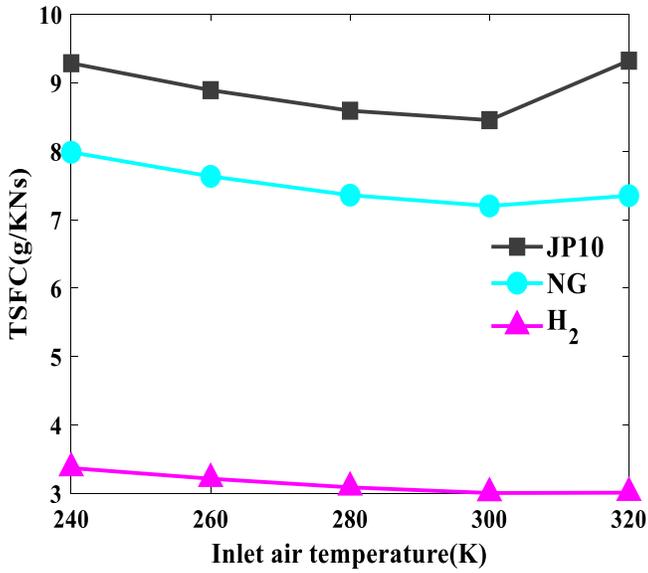 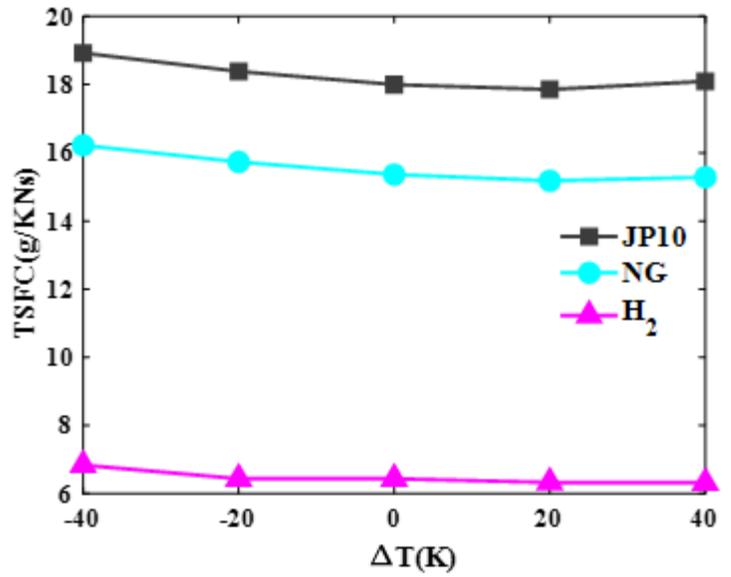

a) b)

**Figure 7.** Specific Fuel consumption changes (TSFC) for input air mass flow rate variations in a) take-off condition and b) on design condition for GENX 1B70, engine.

Figure 8 illustrates changes in snox with inlet air temperature changes according to Kelvin in take-off condition (Ma = 0, H = 0) and on design condition (Ma=0.85, H=10000 m) for GENX 1B70, with JP10 as a fuel engine. It is observed that by reducing the inlet air temperature in constant inlet air pressure, the SNOx parameter is reduced. In this study, the effect of the chemical composition, relative humidity, and combustor geometry on the SNOx was not considered, and only the effect of inlet air temperature & inlet air pressure on the of SNOx was investigated.

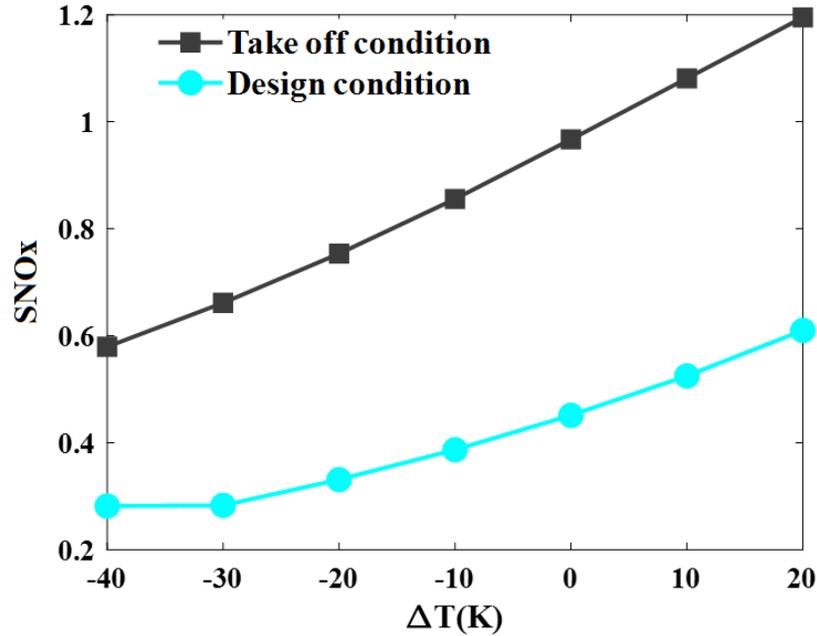

**Figure 8.** Changes in snox with inlet air flow temperature changes according to Kelvin in a) take-off condition and b) on design condition for GENX 1B70, with JP10 as a fuel.

## 3.2. Exergy Analysis

Figure 9 displays exergy destruction rate and exergetic efficiency changes of GENX 1B70 engine subsystems with the inlet air flow temperature at take-off conditions and on design conditions for JP10 fuel. It was observed that in on design conditions, with lessening input air flow temperature, the fan exergetic efficiency, HPC exergetic efficiency, and HPT exergetic efficiency decrease. Also, with lessening the input air flow temperature, the combustor, and the LPT exergetic efficiencies increase. Also, in take-off conditions, it was observed that by reducing the inlet air flow temperature, the exergetic efficiency of the fan, LPT, and HPC increase, also HPT exergetic efficiency decreases.

Also, in on design conditions, it was observed that with decreasing the inlet air flow temperature, the rate of exergy destruction of the fan and HPT increases. Whereas, with decreasing the inlet air flow temperature, exergy destruction rate of the HPC and LPT and combustor is reduced. Also observed in the take-off condition with reducing the inlet air flow temperature, exergy destruction rate of the fan, HPC, and the LPT decreases. By reducing the inlet air flow temperature, the rate of exergy destruction of HPT increases.

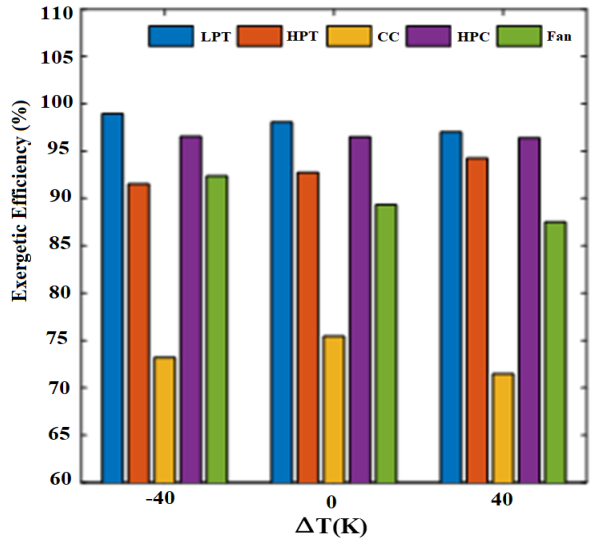
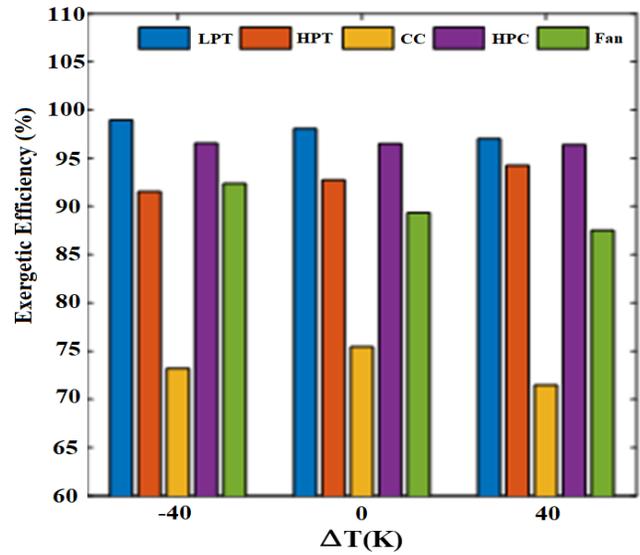

a)

b)

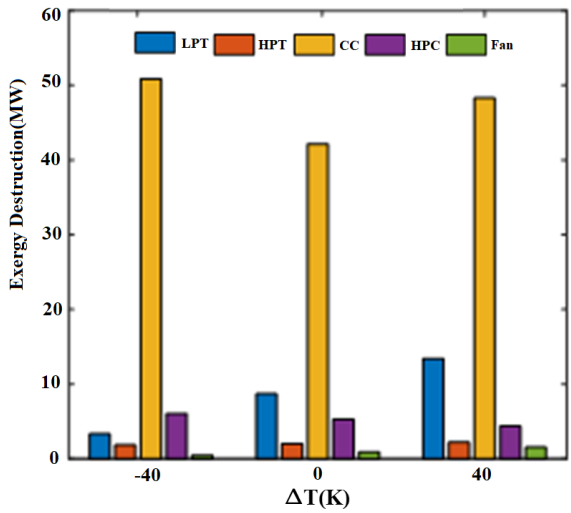
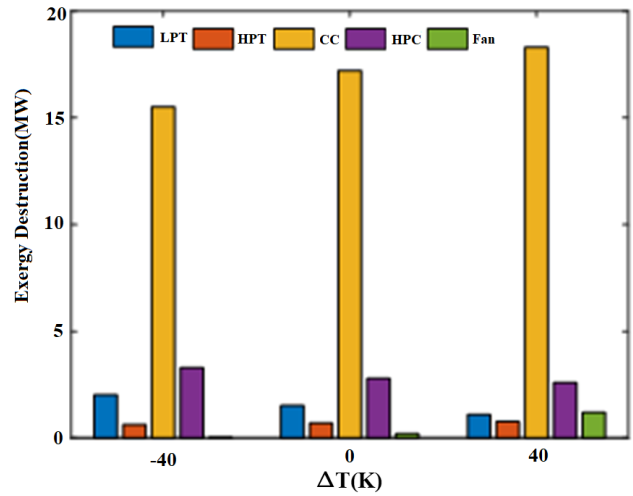

c)

d)

**Figure 9.** Exergetic efficiency and exergy destruction rate of GENX 1B70 engine components changes with the inlet air flow temperature at a) exergy efficiency at the take-off conditions and b) exergy efficiency at the on-design conditions c) exergy destruction at the take-off conditions, and d) exergy destruction at the on-design conditions for JP10 fuel.

Chemical exergy rate of fuel flow and overall exergetic efficiency changes of GENX 1B70 engine with the inlet air temperature at on design conditions for JP10 fuel are shown in Figure 10. With lessening input air flow temperature, the rate of fuel mass flow increases, so the fuel flow chemical exergy rate also increases.

Also, the maximum and minimum rate of chemical exergy flow of fuel flow under the same conditions of among the fuels studied are hydrogen and JP10 fuels, respectively. The overall exergetic efficiency changes of GENX 1B70 engine with the inlet air temperature at on design conditions are shown in Figure 10. As the inlet air flow temperature decreases, the thrust and exergy flow rate of the fuel flow increases. Therefore, by dropping the temperature of the inlet air to the engine, the overall exergetic efficiency decreases. It was also observed that the highest and lowest overall exergetic efficiency of the studied fuels in the same condition belong to JP10 and hydrogen fuel, respectively.

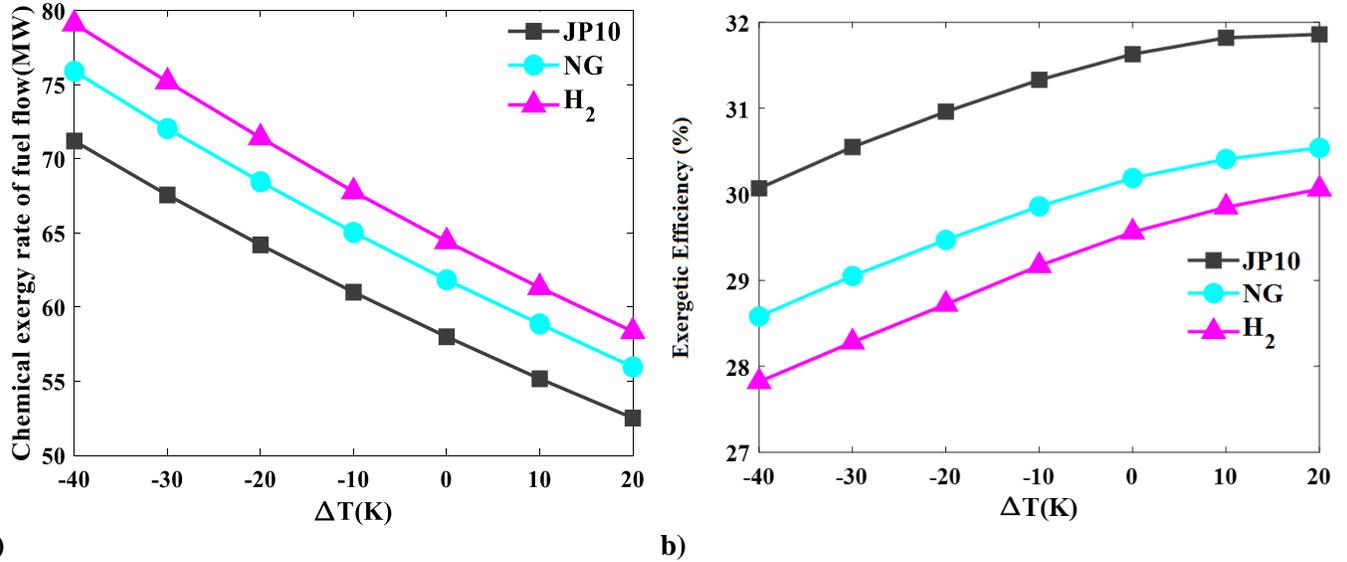

a)  b)

**Figure. 10.** Chemical exergy rate of fuel flow and overall exergetic efficiency changes of GENX 1B70 engine with inlet air flow temperature. a) Chemical exergy rate of fuel flow changes of GENX 1B70 engine and b) overall exergetic efficiency changes of GENX 1B70 engine.

Total entropy generation rate changes of GENX 1B70 engine with the inlet air temperature at on design conditions are shown in Figure 11. As the overall exergetic efficiency of the engine decreases with the inlet air temperature drop, the total rate of entropy generation of the engine increases. It was also observed that the highest and lowest total entropy generation rates between fuels studied in the same condition belong to hydrogen fuel and JP10, respectively.

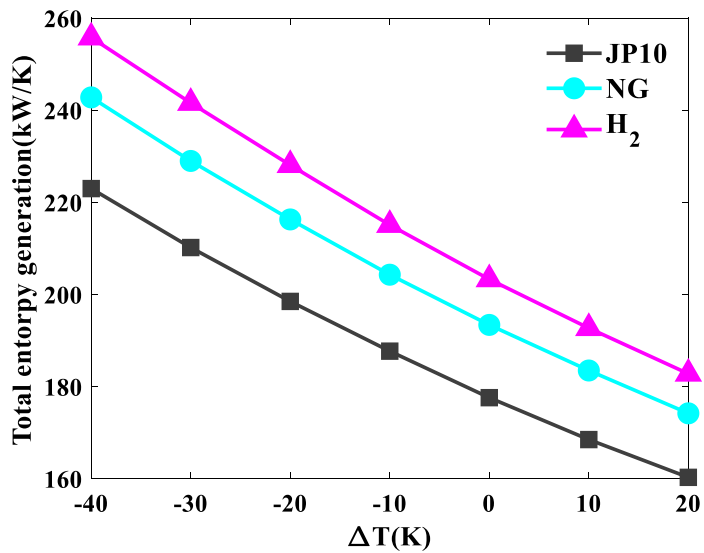

**Figure 11.** Total entropy generation rate changes of GENX 1B70 engine with the inlet air temperature at on design conditions.

### 3.3. Optimization

To optimization GENX 1B70 engine, three separate objective functions were used. Also, a separate optimization was performed for each objective function. These objective functions are: case 1) thrust force maximize, case 2) thermal efficiency maximize, and case 3) propulsive efficiency maximize. In this optimization, TIT, pressure ratio of the compressor, fan pressure ratio, by-pass ratio, and inlet air temperature with ambient air temperature different based on International Standard Atmosphere (ISA) are considered as design variables. In order to reduce the temperature of inlet air to the engine, a cooling compression chiller was employed. It is assumed that the coefficient of performance is 6 for this refrigeration cycle (cooling system). The requisite mechanical power for refrigeration is supplied in the form of off-take power from the engine's high-pressure turbine. According to studies, the thrust force of the thermodynamic cycle is the maximum for liquid hydrogen fuel. It also has the highest thermal efficiency but lowest TSFC compared to other fuels studied in the same condition, so it was selected to optimize hydrogen fuel. The design variables and their ranges are illustrated in Table 4. Optimization constraints based on the objective function for GENX 1B70 engine are given in Table 5. Optimization was performed using a genetic algorithm and according to design variable ranges, optimization constraint, and objective functions and by using a genetic algorithm. The optimal design parameters for GENX 1B70 engine are shown in Table 6 .The results of the optimization have been compared with the results of the baseline engine cycle for hydrogen fuel usage in the on-design condition for GENX 1B70 is specified in Table 7. The results of comparison show that between the optimized cycles and the Base engine cycle, the highest thermal efficiency for case 2 is 0.25% higher than the thermal efficiency of the Base engine cycle. The highest propulsive efficiency and exergitic efficiency and overall efficiency for case 3 are 12.09%, 6.59% and 7.6% respectively, higher than the propulsive efficiency and exergitic efficiency and overall efficiency of the Base engine cycle. Also, the lowest amount of nitrogen oxid products is related to case 3 cycle, which is 42.02% less than the products of Base engine cycle. Also, the highest thrust force and the lowest TSFC are for case 1 cycle, which has increased by 19.43% and decreased by 0.21%, respectively, compared to the thrust and TSFC of the Base engine cycle.

**Table 4. The design variables and their ranges for GENX 1B70 engine.**

| design variables | Turbine inlet temperature(K) | ΔT(K) | Fan compression ratio | Compressor pressure ratio | By-pass ratio |
|---|---|---|---|---|---|
| Lower bounds | 1000 | -10 | 3 | 28 | 7.5 |
| Upper bounds | 2000 | 10 | 4.5 | 32 | 10 |

**Table 5. Optimization constraints based on the objective function for GENX 1B70 engine.**

| Target | Constrains | Minimum value | Maximum value |
|---|---|---|---|
| case 1 | Thermal efficiency | 0.50 | 0.75 |
|  | TSFC(g/KNs) | 2 | 8 |
|  | Propulsive efficiency | 0.75 | 0.95 |
| case 2 | TSF(KNs/Kg) | 140 | 160 |
|  | TSFC(g/KNs) | 2 | 8 |
|  | Propulsive efficiency | 0.75 | 0.95 |
| case 3 | TSF(KNs/Kg) | 70 | 160 |
|  | TSFC(g/KNs) | 2 | 8 |
|  | Thermal efficiency | 0.50 | 0.75 |

**Table 6. The optimal design parameters for GENX 1B70 engine.**

| objective functions | Turbine inlet temperature (K) | ΔT (K) | Fan compression ratio | Compressor pressure ratio | By-pass ratio |
|---|---|---|---|---|---|
| case 1 | 1972.67 | -0.132 | 1.752 | 28.505 | 10.666 |
| case 2 | 1921.05 | -9.954 | 1.432 | 31.998 | 9.008 |
| case 3 | 1494.71 | -9.765 | 1.138 | 32 | 11.998 |

**Table 7. Comparison of GENX 1B70 engine performance in optimal and basic mode for hydrogen fuel.**

|  | case 1 | case 2 | case 3 | Based engine |
|---|---|---|---|---|
| Thrust (kN) | 87.50 | 81.78 | 37.38 | 73.26 |
| Power off-take (kW) | 52 | 932 | 923 | 50 |
| TSFC (g/KNs) | 6.58 | 7.996 | 8.008 | 6.594 |
| Thermal efficiency (%) | 57.91 | 58.06 | 54.85 | 57.91 |
| Propulsive efficiency (%) | 74.74 | 76.01 | 87.38 | 77.95 |
| Overall efficiency (%) | 43.28 | 44.13 | 47.92 | 45.14 |
| Exergetic efficiency (%) | 28.761 | 23.64 | 30.85 | 28.67 |
| Overall entropy generation rate (kJ/K) | 247.3 | 301.50 | 95.60 | 208 |
| Nitrogen oxides generation rate (g/s) | 5.186 | 5.961 | 2.733 | 4.714 |

It can be understood that the changes in on-design conditions of GENX 1B70 engine are different by using the optimal values of the design variables. The GENX 1B70 engine performance is different for each objective function, and the decision is a bit difficult. Using the TOPSIS algorithm, the optimization results are prioritized based on two exero-environmental, and economic approaches in on design conditions Since most of the engine performance is on the design point condition. There are four indicators for choosing the optimal cycle, thrust, thermal efficiency, TSFC, and the index of Nitrogen oxide emission produced in the design point condition (Ma = 0.85, H = 10000m) were considered in this optimization. For all indicators in both approaches, weight coefficients are considered in Table 8.

Table 8. weight coefficients for the environment and economic approaches in TOPSIS algorithm.

| Approaches | $\eta_{th}$ | Exergetic efficiency | TSFC | Overall entropy generation rate | Nitrogen oxides generation rate |
|---|---|---|---|---|---|
| exero-environmental | 0 | 0.99 | 0 | -0.95 | -0.90 |
| Economic | 0.99 | 0 | -0.95 | 0 | 0 |

Table 9. The results of decision-making with the exero-environmental approach and the economic approach.

| Objective functions | Scores in economic approach | Scores in exero-environmental approach |
|---|---|---|
| Thrust maximization(case 1) | 0.81 | 0.21 |
| Thermal efficiency maximization(case 2) | 0.18 | 0.02 |
| Propulsive efficiency maximization(case 3) | 0.05 | 0.77 |

The results of decision-making with the exero-environmental approach and the economic approach are exposed in Table 9. It was detected that in the exero-environmental approach, the highest and lowest scores were obtained by case 2 (propulsive efficiency maximization) and case 2 (thermal efficiency maximization) cycles, respectively. Therefore, the case 2 (propulsive efficiency maximization) cycle was selected by the exero-environmental approach. Base on the economic approach, the highest and lowest score is obtained by case 1 (thrust maximization) and case 2 (propulsive efficiency maximization) cycles, respectively. Therefore, the case 1 (thrust maximization) cycle is chosen base on the economic approach.

## 4. Conclusions

One of the topics of interest for researchers is to increase the power of thrust and reduce fuel consumption and NOx production in turbofans. In this paper, the efficacy of changing the type of fuel and cooling of the intake air on the performance parameters and second law efficiency and the rate of entropy production of the GENX 1B70 engine in both on design and take-off conditions were investigated. One of the distinctive features of this research with other research is turbofan optimization by considering the cooling of the incoming air and selecting the optimal mode based on economic, thermal and environmental criteria using the TOPSIS method. The findings of the present study are summarized:

- In take-off and on design conditions by reducing the temperature of inlet air, the thermal efficiency, thrust force, fuel consumption rate increase, the exergetic efficiency, SNOx, and TSFC decrease. Also, as the temperature of inlet air decreases, the engine entropy production rate increases.
- The results indicate that in the case of using hydrogen fuel, compared to the use of hydrocarbon fuels, the thrust force, thermal efficiency, entropy production rate increase, the exerctic efficiency, and TSFC decrease.
- In the design point conditions, by decreasing the temperature of inlet air flow, the exergetic efficiency of the fan, and HPT decrease, also, the exergetic efficiency of the HPC, LPT, and combustor increases. In take-off conditions, by lessening the input air temperature, the exergetic efficiency of the low-pressure turbine, fan and the high-pressure compressor increase, and the exergetic efficiency of the high-pressure turbine decrease.
- Among the optimal cycles, the highest overall efficiency belongs to the Propulsive efficiency maximization cycle and is equal to 47.92%. The lowest TSFC belongs to the Thrust maximization cycle and is equal to 6.58 g / kNs. Also, the lowest entropy generation rate and the lowest nitrogen oxides production rate belong to the Propulsive efficiency maximization cycle are equal to 95.6 kJ/K and 2.733 g/s, respectively. TOPSIS decision-making method was used to prioritize optimal cycles based on two exero-economic and environmental approaches. The superior cycle, based on the exero-economic point of view and the environmental point of view, is the optimized cycle of maximum thrust and the optimized cycle of maximum propulsive efficiency, respectively.

**AUTHOR INFORMATION**

**Corresponding Authors**

**Somayeh Davoodabadi Farahani** – School of Mechanical Engineering, Arak University of Technology, 38181-41167, Arak, Iran; 0000-0003-1231-1538; Email: sdfarahani@arakut.ac.ir

**Amir Mosavi** – John von Neumann Faculty of Informatics, Obuda University, 1034 Budapest, Hungary; Institute of Information Society, University of Public Service, 1083 Budapest, Hungary; Institute of Information Engineering, Automation and Mathematics, Slovak University of Technology, 812 37 Bratislava, Slovakia; 0000-0003-4842-0613 Email: amir.mosavi@nik.uni-obuda.hu

**Authors**

**Mohammadreza Sabzehali** – School of Mechanical Engineering, Arak University of Technology, 38181-41167, Arak, Iran; 0000-0002-6122-7407; Email:mohamadrezasabzeali19984@gmail.com

**Somayeh Davoodabadi Farahani** – School of Mechanical Engineering, Arak University of Technology, 38181-41167, Arak, Iran; 0000-0003-1231-1538; Email: sdfarahani@arakut.ac.ir

**Amir Mosavi** – John von Neumann Faculty of Informatics, Obuda University, 1034 Budapest, Hungary; Institute of Information Society, University of Public Service, 1083 Budapest, Hungary; Institute of Information Engineering, Automation and Mathematics, Slovak University of Technology, 812 37 Bratislava, Slovakia; 0000-0003-4842-0613 Email: amir.mosavi@nik.uni-obuda.hu

**Nomenclature**

| | |
|---|---|
| $C_{Pavhpt}$ | Specific heat at the HPT(J/kgK) |
| $k_{diff}$ | Ratio of specific heat at the diffuser |
| $A_8$ | Hot stream nozzle area(m$^2$) |
| $A_9$ | Cold stream nozzle area(m$^2$) |
| $C_{Pavlpt}$ | Specific heat at the LPT(J/kgK) |
| $C_{avcc}$ | Specific heat at the combustion chamber(J/KgK) |
| $K_{nh}$ | Ratio of specific heat at the hot path nozzle |
| $K_{nc}$ | Ratio of specific heat at the cold path nozzle |
| $S_{NOx}$ | The nitrate oxide emission index coefficient(g/s) |
| $V_0$ | Flight speed(m/s) |
| $g_c$ | Gravity acceleration (m.s$^{-2}$) |
| $k_{hpt}$ | Fraction of specific heat at the HPT |
| $k_{ch}$ | Fraction of specific heat at the high pressure compressor |
| $k_{cl}$ | Fraction of specific heat at the low pressure compressor |
| $k_f$ | Fraction of specific heat at the fan |
| $k_{lpt}$ | Fraction of specific heat at the LPT |
| $m_a$ | Mass flow rate of the air flow (Kg/s) |
| $m_h$ | Hot stream real air mass flow rate (Kg/s) |
| $m_c$ | Cold stream real air mass flow rate (Kg/s) |
| $m_f$ | fuel mass flow rate(kg/s) |
| $m_t$ | Engine-inlet air real mass flow rate (Kg/s) |
| $\eta_{Ch}$ | High pressure compressor isentropic efficiency |
| $\eta_{Cl}$ | Low pressure compressor isentropic efficiency |
| $\eta_{HPT}$ | High pressure turbine isentropic efficiency |
| $\eta_{LPT}$ | Low pressure turbine isentropic efficiency |
| $\eta_{combustor}$ | Combustor isentropic efficiency |
| $\eta_{fan}$ | Fan isentropic efficiency |
| $\eta_{nh}$ | Hot stream nozzle isentropic efficiency |
| $\eta_{nc}$ | Cold stream nozzle isentropic efficiency |
| $\eta_o$ | overall efficiency |
| $\eta_p$ | propulsive efficiency |
| $\eta_{th}$ | thermal efficiency |
| $\pi_{Ch}$ | High-pressure compressor pressure ratio |
| $\pi_{Cl}$ | Low-pressure compressor pressure ratio |
| $\pi_{fan}$ | Fan pressure ratio |
| A | cross-section of an air inlet flow(m$^2$) |
| HPC | High pressure compressor |
| HPT | High pressure turbine |
| LPC | Low pressure compressor |
| LPT | Low pressure turbine |
| Ma | Flight Mach number |
| H | Flight altitude |
| R | Global constant gas(J/kgK) |
| TIT | High-pressure turbine inlet temperature(K) |
| TSF | Specific thrust(g/KNs) |
| $FHV$ | Fuel heat value (MJ/Kg) |
| $TSFC$ | Especial fuel consumption(g/KNs) |
| $\alpha$ | By-pass ratio |
| $\rho$ | Inlet air density (Kg.m$^{-3}$) |

**Abbreviation**

| | |
|---|---|
| SNOx | Nitrogen oxide emission intensity index |
| TOPSIS | Technique for Order of Preference by Similarity to Ideal Solution |

**For Table of Contents Only**

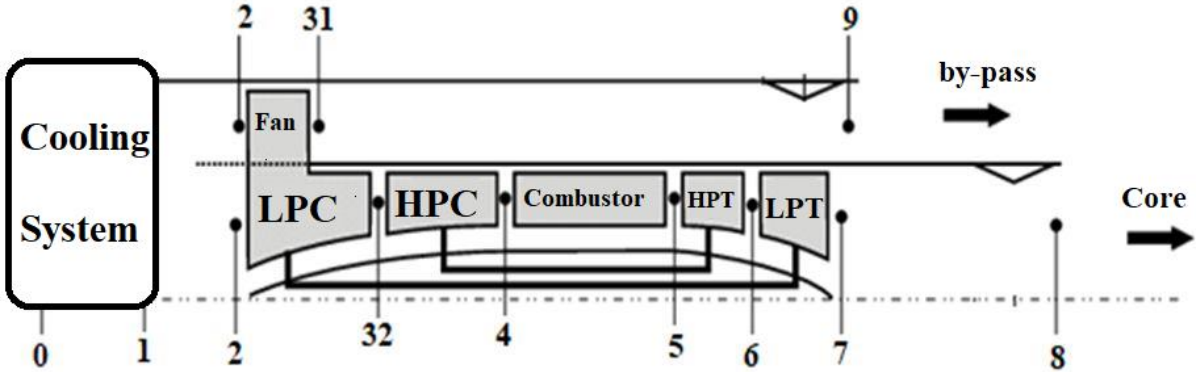